\documentclass[showpacs,preprintnumbers,amsmath,amssymb]{revtex4}
\usepackage{graphicx}
\usepackage{dcolumn}
\usepackage{bm}
\begin{document}
\title{Unparticle may be a Remedy to the  Information Loss in the Scattering of Fermion
off Dilaton Black-hole}
\author{Anisur Rahaman} \email{1. anisur.rahman@saha.ac.in, 2.
manisurn@gmail.com} \affiliation{Hooghly Moghin College,
Chinsurah, Hooghly-712101, West Bengal, India}

\date{\today}

\begin{abstract}
We study s-wave scattering of fermion off dilaton black-hole. With
one loop correction it was found to suffer from nonpreservation of
information and that of course, went against Hawking's revised
suggestion on this issue. A nonstandard approach, e.g. the
probable existence of unparticle in $(1+1)$ dimension has been
adopted here that shows a remedy to get rid of the danger of
information loss to bring it in  agrees with the Hawking's revised
suggestion.
\end{abstract}

 \maketitle
\section{Introduction}
 There has been much interest in the physics of  back-hole over
 the last few decades and the scattering of
fermion off dilaton black hole carries special interest because
the literatures related to this problem \cite{STRO, CALL, SUS,
GARF, GIDD, TBANK,REV,VEG, MIT, AR, AR1} have provided much
insight into its connection to the Hawking radiation. The issue
that attracted attention to a great extent was the possible
information loss during the formation and the subsequent
evaporation of the black-hole. A controversy in this context was
generated from Hawking's initial suggestion about four decades ago
\cite{HAW}, but his revised suggestion on this issue \cite{HAW1}
has brought back moderately pleasant scenario. In spite of that,
it is fair to admit that the matter related to the preservation of
information did not yet conclusively settl down from all corner.
Therefore, investigation related to this problem still be of
interest.

A general description of this process was pursued in \cite{HOOFT},
where it was found to be formulated  through the s-matrix
description of the event which involved lots of inherent
constructional complexity and computational difficulty. Therefore,
most of the authors preferred to handle this problem with less
complicated but interacting and well formulated model \cite{STRO};
see \cite {REV} for review. The model took birth in the
description of two dimensional noncritical string theory and the
black hole solution of this was carried out in \cite{GAS}. An
extremal magnetic charged black-hole solution was  considered in
these studies. Eventually, it was generated from  a (3+1)
dimensional model involving a dilaton field. Since in the study of
s-wave scattering of fermion in this black-hole, angular
coordinates becomes irrelevant and an interesting (1+1)
dimensional effective action takes birth. It is known from the
important publication \cite{STRO}, that when the energy involved
in this process is not too high, the metric and the dilaton field
can be treated as external classical quantities and an amusing
version of quantum electrodynamics with position dependent
coupling constant emerges out.

With this framework the scattering of massless Dirac fermion was
studied in \cite{STRO}. Although the mathematical formulation did
involve a general description (as permitted within this
framework), the authors did not encounter the problem of
information loss since one loop correction was not included there.
However, with the similar framework, when this scattering
phenomena was extended with the chiral fermion in \cite{MIT} the
author had to face the danger of real information loss. The one
loop correction \cite{JR, KH} entered in this description
automatically during bosonization. The present author in
\cite{ARINF}, showed that there exists a remedy of this danger if
the possibility of allowance of anomaly is exploited judiciously.
Note that, this framework is so powerful and beautifully designed
that it itself has a room for taking the anomaly into
consideration. In the study of scattering of Dirac fermion
\cite{STRO}, though the author did not face the danger of
information loss, the present author reported that information
loss could not be avoided when one loop correction that entered
during the process of bosonization was taken into account
\cite{AR}. The result though led to an uncomfortable situation
there was no known standard physical principal to avoid it as it
was found in the case of chiral fermion \cite{ARINF}. It certainly
diminishes the glory of this important and well formulated
framework of studying the scattering problem \cite{STRO} since the
result went against Hawking's revised suggestion \cite{HAW1}. In
this work, therefore, an attempt has been made to bring it in
agreement with Hawking's revised suggestion \cite{HAW1} making it
free from the danger of information loss even in the presence of
one loop correction.

A failure to find the standard physical principal towards having a
plausible solution of a standing unresolved problem does not mean
that some non standard approach would not not be of worth for
that. A very good instance in this direction is the introduction
of unparticle to get a plausible solution of momentum distribution
\cite{GEOR} at high energy regime. After the introduction of the
concept of unparticle  in the seminal work \cite{GEOR}, it has
been used in different important studies. The studies related to
dark matter using this concept is of worth mentioning \cite{KIKU,
DARK}. Study of Casimir effect in presence of unparticle is
another important phenomenological development \cite{CASI}. An
attempt to establish unparticle  as holographic dual of gapped AdS
gravity is also an interesting theoretical improvement
\cite{SOPH}. Unparticle with one loop correction was studied in
\cite{ARUN} where a remarkable retrieval of gauge invariance in
the usual phase space was found. In this work we are, therefore,
intended to exploit the advantage of having the probable existence
of unparticle in (1+1) dimension \cite{GEOR1} to get a remedy of
the problem of information loss that occurred in the s-wave
scattering process when one loop correction got involved in
\cite{AR}. Let us mow turn towards that end.

The Paper is organized as follows. Sec. II,  is devoted with a
brief discussion of Unparticle in $(1+1)$ dimension to make the
paper self contained. Sec. III deals with the formulation of the
model related to s-wave scattering of fermion off dilaton black
hole that is expected to give a plausible solation of information
loss. In Sec. IV, we determine the theoretical spectrum of the
model using Dirac's prescription of quantization of constrained
system. Sec. V, contains a description of the plausible solution
towards removal of information loss problem setting a cardinal
scale and the final Sec. VI, contains the concluding remarks.

\section{A brief description of Unparticle in two dimension}
To make this article self contained it would be useful to start
with a brief introduction of unparticle, to be precise unparticle
in $(1+1)$ dimension, before going to the actual formulation of
the model which is aimed to get a plausible remedy of the
information loss in the Scattering of Fermion off Dilaton
Black-hole.

We are habituated to see our  quantum mechanical world in terms of
particles. In the classical domain these a particle has definite
mass and therefore carry energy and momentum in a definite
relation $E^2 = p^2c^2 + m^2c^4$, which turns in to a dispersion
relation in quantum mechanics \cite{GEOR}. The scale invariance
can not be seen unless that mass reduces to zero. Therefore, a
free massless particle is a very good and simple example of scale
invariant stuff because the zero mass remains unaffected by
resealing. The standard model though does not have the property of
scale invariance but there could be a sector of the theory, which
is yet unseen but it is exactly scale invariant and very weakly
interacting with the rest of the standard model.

This reminds the work of Banks and Zaks \cite{BANKS} where they
observed a non-trivial zero of the $\beta$ function in the IR
region of Yang-Mills theories with certain non-integral number of
fermions that indicates the absence of a particle like
interpretation. Georgi in his seminal work \cite{GEOR}, termed
this scale invariant stuff as unparticle and formulated an
attactive frame work. The topic attracted a huge attention and a
variety of fields of research, spanning astrophysics, neutrino
physics, AdS/CFT duality and quantum gravity have found its
application.

The exactly soluble 2D theory of a massless fermion coupled to a
massive vector boson, the Sommerfield  model (Thirring-Wess), is
an interesting analog of a Banks-Zaks model. It is approaching to
a free theory at high energies and a scale invariant theory with
nontrivial anomalous dimensions at low energies. This is a toy
standard interacting model where interaction of fermions with the
gauge field is considered which shows a transition from unparticle
behavior at low energies to free particle behavior at high
energies.

The Sommerfield (Thirring-Wess) model model \cite{SOM,THIR} is
described by the action
\begin{equation}
{\cal S}_f = \int d^2x[i\bar\psi\gamma^\mu(\partial_\mu -
ieA_\mu)\psi - \frac{1}{4}F_{\mu\nu}F^{\mu\nu}]. \label{THW}
\end{equation}
Here $e$ has one mass dimension representing the coupling constant
between the vector current associated with the massless fermion
field $\psi$ and the gauge field $A_\mu$. The Lorentz indices
$\mu$ and $\nu$ takes the values $0$ and $1$ corresponding to a
$(1+1)$ dimensional space time and $F_{\mu\nu}$ is the
electromagnetic field strength. It is the well known vector
Schwinger model \cite{SCHW} with Proca background, i.e., Schwinger
model with an additional masslike term for the gauge field. Both
the Schwinger model and the Sommerfield (Thirring-Wess) model are
exactly solvable model where gauge field interacts vectorially
with the current associated with the fermion field. The
theoretical spectrum of the Schwinger model shows only a massive
particle with mass $\tilde{e}$, where $\tilde{e} =
\frac{e}{\sqrt{\pi}}$, however the solution of Sommerfield
(Thirring-Wess) model shows the presence of a massless along with
a  massive field with square of the mass $m^2 = m_0^2 +
\tilde{e}^2$. Alhough the Schwinger model does not have a scale
invariant sector the Sommerfield (Thirring-Wess) model does have.
A structurally identical model (in algebraic sense), the so called
the Nonconfinig Schwinger model was studied by us in \cite{PAR}
where the masslike term for gauge field  although
 got involved in completely different perspective may certainly be
useful to understand the the theoretical spectra in the present
situation. In fact, that term was entered there through one loop
correction in order to remove the divergence of the fermionic
determinant appeared during bosonization. The contents of the
article \cite{PAR} is able to provide algebraically a clear
exposition of the solution of Sommerfield (Thirring-Wess) model
trough the determination of theoretical spectrum using constrained
dynamics due to Dirac \cite{DIR}. It can be seen just by setting
$m_0^2 =\frac{ag^2}{\pi}$ in  \cite{PAR}. It is also shown there
that the gauge field propagator shows the poles at the expect
positions \cite{PAR}. To get a clear view, it would be of worth to
write down the effective action that follows from (\ref{THW}) in
the manner it has been computed in \cite{PAR} :
\begin{equation}
S_{eff}=\frac{1}{2}\int d^2x[A_\mu(x)
M^{\mu\nu}A_\nu(x)],\end{equation} where,
\begin{equation}
M^{\mu\nu}=m_0^2g^{\mu\nu} -
\frac{\Box+\tilde{e}^2}{\Box}\tilde\partial^\mu\tilde\partial^\nu.\end{equation}
Here the standard notation
\begin{equation}
\tilde\partial^\mu=\epsilon^{\mu\nu}\partial_\nu.
\end{equation}
has been used.  The antisymmetric tensor is defined with the
convention $\epsilon^{01}=-1$. From the the effective action, the
gauge field propagator is found to be
\begin{equation}
\Delta_{\mu\nu}(x-y)=\frac{1}{m_0^2}[g_{\mu\nu}+ \frac{\Box
+\tilde{e}^2}{\Box(\Box + m_0^2 +\tilde{e}^2)}
\tilde\partial_\mu\tilde\partial_\nu]\delta(x-y).
\end{equation}
It shows the two poles at the expected positions: one at vanishing
mass and the other one at $m^2 = m_0^2 + \tilde{e}^2$ \cite{PAR}
From the infirmations available in \cite{GEOR}, it is natural to
accept that physical mass $m= \sqrt{m_0^2 +\tilde{e}^2}$ plays the
role of unparticle scale $\Lambda_m$ of this in $(1+1)$
dimensional model. Here we find the presence of a free massless
scale invariant sector. Georgi himself has identified this
massless sector of the theory (\ref{THW}) as unparticle
\cite{GEOR1} and the mass of the massive scaler was pointed out
there as unparticle scale in \cite{GEOR1} which agrees with his
previous seminal work\cite{GEOR}. Let us now turn towards the
actual formulation of the model which will serve our present
purpose.

\section{Formulation of the model towards getting a remedy of
 the Information loss}
The model for studying the scattering of fermion off dilaton black
hole was formulated with an interacting Dirac fermion in a dilaton
back ground \cite{STRO, SUS}. The lagrangian density that
describes it is contained within the action
\begin{equation}
{\cal S}_f = \int d^2x[i\bar\psi\gamma^\mu[\partial_\mu -
ieA_\mu]\psi - \frac{1}{4} e^{-2\Phi(x)}F_{\mu\nu}F^{\mu\nu}].
\label{EQ1}
\end{equation}
 The dilaton field $\Phi$ stands as
a non dynamical back ground. This model is considered here to
study the s-wave scattering of fermion off dilaton black hole. It
is known that the black hole is the extrema of the following (3+1)
dimensional action
\begin{equation}
S_{AF} = \int d^4 x\sqrt{-g}[R + 4(\nabla\phi)^2 - \frac{1}{2}F^2
+ i \bar\psi D\!\!\!/\psi].
\end{equation}
Here $g$ stands for the determinant of the metric $g_{\mu\nu}$,
$\psi$ is the charged fermion, $D_\mu = \partial_\mu - ieA_\mu$,
$F^2= F_{\mu\nu}F^{\mu\nu}$ and $R$ is the scalar curvature.

The geometry involved here consists of three region \cite{ STRO,
REV}. Far from the black hole there is an asymptotically flat
region. At the close vicinity the curvature begins to rise. It is
essentially a mouth region. The mouth leads to an infinitely long
throat. Inside the throat the metric is approximated by a flat
metric on two dimensional Minkowsky space times the round metric
on two sphere with radius $Q$. Deep into the throat the low energy
physics is effectively two dimensional and it remains confined in
a plane. The two dimensional effective theory can be viewed as a
compactified form of four to two dimension and the standard
Kaluza-Klein technique leads to the effective $(1+1)$ dimensional
action \cite{STRO}
\begin{equation}
S_{AF} = \int d^2\sigma\sqrt{g}[R + 4(\nabla\phi)^2 + \frac {1}
{Q^2} - \frac{1}{2}F^2 + i \bar\psi D\!\!\!/\psi] \label{TWA}
\end{equation}
For sufficiently low energy incoming fermion gravitational effect
can be neglected and equation (\ref{TWA}) reduces to equation
(\ref{EQ1})

To bring the unparticle into action in this scattering phenomena
we modify the action (\ref{EQ1}) with an additional masslike term
for gauge field following the line of action of Sommerfield
(Thirring-Wess) model.
\begin{equation}
{\cal S}_f = \int d^2x[i\bar\psi\gamma^\mu[\partial_\mu -
ieA_\mu]\psi - \frac{1}{4} e^{-2\Phi(x)}F_{\mu\nu}F^{\mu\nu} +
\frac{1}{2}m_0^2 A_\mu A^\mu]. \label{UNEQ}
\end{equation}
Without violating any physical principle a mass like term for the
gauge field is added here which implies a shifting of the
electromagnetic background from Maxwell field to Proca and a
surprisingly entry of the unparticle results in. We refer the work
of Georgi \cite{GEOR1} for the technical details of the probable
appearance of unparticle in lower dimensional physics. In the
absence of one loop correction  the theoretical spectra of the
above model (\ref{UNEQ}) shows to have a massless scalar along
with a massive scalar  having square of the mass mass $m^2=m_0^2 +
e^{2\Phi(x)}e^2$. To see how the factor $e^{2\Phi(x)}e^2$ enters
into the mass term needs detailed constraint analysis and
determination of theoretical spectrum of the model (\ref{UNEQ})
but it can be understood from our previous work \cite{AR} and the
earlier work in this context reported in \cite{MIT}. The physical
mass $m= \sqrt{m_0^2 + e^{2\Phi(x)}e^2}$ as usual plays the role
of unparticle scale \cite{GEOR, GEOR1}. It sets the scale of the
transition from free particle behavior at high energy to the
unparticle behavior at low energy. This unparticle scale will be
modified if we include the one loop correction as we did in
\cite{PAR}. The one loop correction here too enters automatically
during the process of bosonization \cite{PAR}. Here bosonization
is done integrating the fermion out one by one which leads to a
determinant carrying a singularity. To remove that singularity it
is needed to regularize it \cite{PAR}. After proper
regularization, when the determinant is written in terms of
auxiliary field the following lagrangian density results.
\begin{equation}
{\cal L}_B = \int d^2x[\frac{1}{2}\partial_\mu \phi
\partial^\mu \phi - \tilde{e} \epsilon_{\mu\nu}\partial^\nu\phi A^\mu
- \frac{1}{4} e^{-2\Phi(x)}F_{\mu\nu}F^{\mu\nu} +
\frac{1}{2}\tilde{e}^2(a+\frac{m_0^2}{\tilde{e}^2}) A_\mu A^\mu].
\label{LAGB}
\end{equation}
The parameter $a$ is the ambiguity parameter of regularization
with which we are familiar from quite a long past \cite{JR, KH,
PAR}.

\section{Quantization of the model to get theoretical spectrum}
To get the theoretical spectrum in the present situation, let us
now proceed with the constraint analysis of the resulting theory
described by the lagrangian density (\ref{LAGB}). It necessities
the computation of the momenta corresponding to the fields
describing the Lagrangian (\ref{LAGB}). The canonical momenta
corresponding to the scalar field $\phi$, the gauge field $A_0$
and $A_1$ as obtained from the standard definition are
\begin{equation}
\pi_\phi = \dot\phi + \tilde{e}A_1,\label{MO1}
\end{equation}
\begin{equation}
\pi_0 = 0,\label{MO2}
\end{equation}
\begin{equation}
\pi_1 = e^{-2\Phi(x)}(\dot A_1 - A_0') \label{MO3}
\end{equation}
Here $\pi_\phi$, $\pi_0$ and $\pi_1$ are the momenta corresponding
to the field $\phi$, $A_0$ and $A_1$. Using the above equations
(\ref{MO1}), (\ref{MO2}) and (\ref{MO3}), it is straightforward to
obtain the canonical Hamiltonian through a Legendre
transformation. The canonical Hamiltonian is found out to be
\begin{equation}
H_C = \int dx^1[\frac{1}{2}(\pi_\phi + \tilde{e}A_1)^2+
\frac{1}{2} e^{2\Phi}\pi_1^2 + \pi_1A_0' + \frac{1}{2}\phi'^2 -
\tilde{e}A_0 \phi' -\frac {1}{2}\tilde{e}^2(a+
\frac{m_0^2}{\tilde{e}^2})(A_0^2 - A_1^2].\label{CHAM}
\end{equation}
Note that equation (\ref{MO2}) is a primary constraint of the
theory. The preservation of this constraint (\ref{MO2}) leads to a
secondary constraint
\begin{equation}
\pi_1' +\tilde{e}\phi' + \tilde{e}^2(a+
\frac{m_0^2}{\tilde{e}^2})A_0 \approx 0. \label{SCON}
\end{equation}
The constraints standing in the equations (\ref{MO2}) and
(\ref{SCON}), form a second class set. The system does not have
any other constraint excepting these two. In order to get the
reduced Hamiltonian (according to Dirac's terminology ) we need to
impose these two constraints into the Hamiltonian which ultimately
renders the following Hamiltonian in the reduced phase space.
\begin{eqnarray}
H_R &=& \int dx^1[ \frac{1}{2}(\pi_\phi + \tilde{e}A_1)^2+
\frac{1}{2}\frac{1}{a\tilde{e}^2+m_0^2}(\pi_1' + \tilde{e}\phi')^2
+\frac{1}{2}e^{2\Phi(x)}\pi_1^2 + \frac{1}{2}\phi'^2 \nonumber \\
&+& \frac{1}{2} \tilde{e}^2(a+ \frac{m_0^2}{\tilde{e}^2}) A_1^2].
\label{RHAM}
\end{eqnarray}
The Dirac brackets \cite{DIR} between the fields describing the
reduced Hamiltonian (\ref{RHAM}) remain canonical and using that
canonical set of Dirac brackets we get the following equations of
motion.
\begin{equation}
\dot A_1 = e^{2\Phi(x)}\pi_1 - \frac{1}{a\tilde{e}^2+
m_0^2}(\pi_1'' + \tilde{e}\phi''), \label{EQM1}
\end{equation}
\begin{equation}
\dot\phi = \pi_\phi + \tilde{e}A_1, \label{EQM2}
\end{equation}
\begin{equation}
\dot \pi_1 = -\tilde{e}\pi_\phi - (a\tilde{e}^2+ m_0^2) A_1.
\label{EQM3}
\end{equation}
\begin{equation}
\dot\pi_\phi = \frac{\tilde{e}^2(a+1)+ m_0^2}{a\tilde{e}^2+
m_0^2}\phi'' + \frac{\tilde{e}}{a\tilde{e}^2+m_0^2 }\pi_1''.
\label{EQM4}
\end{equation}
 The above four equations (\ref{EQM1}), (\ref{EQM2}),
(\ref{EQM3}) and (\ref{EQM4}), lead to the following two second
order differential equations after a little algebra.
\begin{equation}
(\Box+\tilde{e}^2e^{2\phi}(a+ \frac{
m_0^2}{\tilde{e}^2}+1)\pi_1=0,
\end{equation}
\begin{equation}
\Box[\pi_1 + \tilde{e}(a+ \frac{ m_0^2}{\tilde{e}^2}+1)\phi]=0.
\end{equation}
Here $\pi_1$ represents a massive boson with square of the mass
$m^2=\tilde{e}^2e^{2\phi}(a+ \frac{ m_0^2}{\tilde{e}^2}+1)$ and
$\phi$ represents a massless scalar field which was termed as
unparticle in \cite{GEOR1}. Note that the physical mass term $m =
\sqrt{\tilde{e}^2e^{2\phi(x)}(a+ \frac{ m_0^2}{\tilde{e}^2}+1)}$
has modified getting the effect of one loop correction which is
playing the role of unparticle scale in the present situation. The
model considered here has position dependent coupling constant
that arose from (3+1) dimensional model involving a dilaton field
and in this situation it is natural that unparticle scale
(physical mass term) would be space dependent but that will not
pose any problem to set the unparticle scale at at a desired value
at a particular space-time point which in fact we need for our
purpose. We will come to this point in detail in the next Sec. The
surprising aspect of this model, i.e., the presence of the space
dependent exponential factor $e^{2\Phi}(x)$, where $\Phi= -x^1$
for the background generated by linear dilaton vacua in $(1+1)$
dimensional gravity makes it amenable to study the s-wave
scattering where information puzzle gets in. The mass of the
massive boson therefore goes on increasing indefinitely towards
the negative $x^1$. So a final contribution will be totally
reflected and an observer at $x^1 \rightarrow \infty$ will get
back all the information. To be more precise mass will vanish near
the mouth but increases indefinitely as one goes into the throat
because of the variation of this space dependent factor $\Phi$.
Since a massless scalar is equivalent to a massless fermion in
$(1+1)$ dimension, this can be thought of as a massless fermion
proceeding into the black-hole it will not be able to travel an
arbitrary long distance and will be reflected back with a unit
probability. Thus there will be no information loss from the
massive sector of the theory. But  a genuine problem will arise
from the massless sector because the massless boson (fermion) will
remain massless irrespective of its position. So it will go on
travelling through out the black hole without any hindrance.
Therefore, an observer at $x^1 \rightarrow \infty $ will never
find this massless fermion with a return journey. So a real
problem related to information loss appears due to the presence of
the massless fermion. It is certainly an unwanted scenario and it
goes against Hawking's revised Suggestion \cite{HAW1}.
Unfortunately, this possibility can't be ruled out if one has to
accept the said model \cite{STRO}. The aim of this work is to find
out a plausible remedy of this problem towards which we now turn

\section{Plausible solution to the remedy of the Information loss}
 To get out of this problem an attempt has been made to adopt a non
 standard approach because till now there is no known standard approach
 to achieve it. To this end we bring the
unparticle into action in order to get a remedial measure. Let us
see how the unparticle has been brought into service to save this
particular problem from the unavoidable information information
loss in presence of one loop correction. We have seen that this
theory with one loop correction contains an ambiguity parameter
$a$. We are free to exploit this ambiguity to set the unparticle
scale $\tilde{\Lambda}_m = \tilde{m}= \tilde{e}\times
e^{\Phi(\tilde{x}^1)}$ at a particular space time point say for
example at $x^1=\tilde{x}^1$ and $x^0=\tilde{x}^0$ , at the throat
region \cite{STRO}. The unparticle scale at $x^1=\tilde{x}^1$ and
$x^0=\tilde{x}^0$, therefore, represents the {\it cardinal} point
of transition from unparticle behavior at low energies to
pertubative behavior at high energies. This setting of scale
immediately brings a drastic change into the constraint structure
of the theory for all space-time points after $x^1=\tilde{x}^1$
and $x^0=\tilde{x}^0$ and that brings a miraculous change into the
theoretical spectrum. In fact, the constraint (\ref{SCON}) gets
modified to
\begin{equation}
\pi_1' +\tilde{e}\phi' \approx 0, \label{MSCON}
\end{equation}
since that setting demands $a+\frac{m_0^2}{e^2}=0$, which can be
met up exploiting the regularization ambiguity. It is amazing to
note that the constraints (\ref{MO2}) and (\ref{MSCON}) turns into
a first class set loosing their second class nature with this
setting and that brings a radical change in the theoretical
spectrum. To get the theoretical spectra at this stage we need two
gauge fixing condition corresponding to the two first class
constraints. The gauge fixing conditions are $A_0=0$ and $A'_1=0$
are now introduced. The canonical Hamiltonian along with the two
constraints (\ref{MO2}), and (\ref{MSCON}) and the above two gauge
fixing conditions reduces to the following simplified form.
\begin{equation}
H_R = \int dx^1[ \frac{1}{2}\pi_\phi^2+ \frac{1}{2} \phi'^2 +
\frac{1}{2} \tilde{e}^2 e^{2\Phi} \phi^2]. \label{MRHAM}
\end{equation}
The Dirac brackets \cite{DIR} in this situation are also found to
remain canonical. These canonical Dirac brackets lead to two first
order differential equation from the Hamiltonian (\ref{MSCON}).
\begin{equation}
\dot\phi=\pi_\phi, \label{EM1}
\end{equation}
\begin{equation}
{\dot \pi}_\phi= \phi'' - \tilde{e}^2e^{2\Phi}\phi. \label{EM2}
\end{equation}
The above two first order differential equation (\ref{EM1}) and
(\ref{EM2}) after a little algebra gets simplified into
\begin{equation}
(\Box+\tilde{e}^2e^{2\phi})\phi=0.\label{SPEC}
\end{equation}
The equation (\ref{SPEC}) describes the theoretical spectrum at
this stage which suggests that there is only a massive boson with
square of the mass $m^2=\tilde{e}^2e^{2\Phi}$. Therefore, unlike
the previous situation, after the space-time point, the system
does not contain any massless boson (equivalent to fermion) when
the unparticle scale is set to $\tilde{\Lambda}_m = \tilde{m}=
\tilde{e}\times e^{\Phi(\tilde{x}^1)}$ at $x^1=\tilde{x}^1$ and
$x^0=\tilde{x}^0$. This setting miraculously eradicates the
massless sector from the system. It was known that the source of
all disastrous related to information loss was laid in the
massless sector of the theory. The introduction of the concept of
unparticle along with the setting of a cardinal scale of energy by
exploiting the ambiguity parameter judiciously the massless sector
of the theory has been found to get eradicated and that ultimately
make it free from the undesired as well as uncomfortable
information loss problem.  This indeed agrees with Hawking's
revised proposal regarding this issue. It also agrees with the
more recent result reported in \cite{VEG1}. We should mention that
the system does not contain any unparticle after setting that
unparticle scale to $\tilde{\Lambda}_m = \tilde{m}=
\tilde{e}\times e^{\Phi(\tilde{x}^1)}$ at $x^1=\tilde{x}^1$ and
$x^0=\tilde{x}^0$. Rendering its great service it gets disappeared
from the spectrum.

A question may be raised: what does actually happen here
physically? The answer lies in the setting of a suitable cardinal
scale of energy which is accessible from the exploitation of the
ambiguity in the one loop term in suitable manner. This scale as
stared earlier represents the cardinal point which indicates the
starting point to transit from unparticle behavior at low energies
to perturbative behavior at high energies. Therefore, from that
point the combination of unparticle and the standard matter field
can be considered as standard matter field only \cite{GEOR1}. So
the combined system effectively turns into a single one which
indeed is  free from the masslees excitation and that ultimately
lead to a information preserving process.

\section{Conclusion}
The model used here is a (1+1)dimensional toy model and all the
finite details of the real black-hole is not contained in it.
However, the information loss issue analogous to the one as raised
by Hawking is contained in it in a significant manner. The
plausible remedy that is obtained from this work although has been
achieved from a nonstandard approach is a novel one, and that of
course, did not violate any physical principle, and above the all
that agrees with the Hawking's revised suggestion on the issue of
information loss \cite{HAW1}. However, we should  admit that we
have to be satisfied with the logical consistency keeping
particular view on the crucial fact such that no physical
principle gets violated since experimental detection  of
unparticle is a remote possibility with the hitherto developed
experimental facilities in the high energy regime. Therefore, this
work adds a novel way of obtaining a remedy to the information
loss without violating any physical principle that ought to be
faced during the s-wave scattering off dilaton black hole when it
is studied with one loop correction through the model described in
(\ref{EQ1}). It indeed bring back the lost glory of the framework
along with the results that is in good agreement with Hawking's
revised suggestion. Finally, the the point on which we would like
to emphasize is that the cardinal scale which represents the
starting point of transition from unparticle behavior at low
energies to perturbative behavior at high energies is a fixed
valued energy although the unparticle scale that results from the
model which has been used for studying s-wave scattering problem
is space dependent.

Setting of unparticle scale to $\tilde{\Lambda}_m = \tilde{m}=
\tilde{e}\times e^{\Phi(\tilde{x}^1)}$ at $x^1=\tilde{x}^1$ and
$x^0=\tilde{x}^0$ needs the requirement $a+\frac{m_0^2}{e^2}=0$. A
careful look reveals that it has a deeper meaning. In fact, an
automatic cancellation of a classical masslike term takes place by
a quantum mechanically generated masslike term. It has made
possible through the very exploitation of ambiguity often found in
the one loop correction that appears during bosonization in order
to remove the singularity in the ferneonic determinant. It can not
be considered as a  simple abolition of the masslike terms
involved in the model. For example a naive choice $a=0$ and
$m_0=0$ may satisfy the requirement $a+\frac{m_0^2}{e^2}=0$.

We are familiar with the exploitation of ambiguity from quite a
long past. The ambiguity which are often found present in the one
loop correction term has been found to exploit several times  to
obtain a fruitful and long standing suffering removal result.
Quite a good number of examples are available in the literature
concerning this issue \cite{MIT, AR1,JR, KH, ARINF, ARUN, ARAN,
ARNL, AAP}. So it is more or less a standard approach now. The
first and famous instance in this context reminds us the removal
of long suffering of chiral Schwinger model from the non unitarity
problem \cite{JR}.

\section{Acknowledgement}
It is my pleasure to thankfully acknowledge the provision of
getting computer facilities of Saha Institute of Nuclear Physics.

\end{document}